\documentclass[aps,prb,reprint,showpacs,amsmath,amssymb,groupaddress,citeautoscript]{revtex4-1}
\usepackage{graphicx}
\usepackage{epstopdf}
\usepackage{color}

\begin{document}

\preprint{E. Lengyel et al.}

\title{Temperature - pressure phase diagram of CeCoSi:\\ Pressure induced high-temperature phase}

\author{E. Lengyel}
\email{lengyel@cpfs.mpg.de}

\author{M. Nicklas}

\author{N. Caroca-Canales}

\author{C. Geibel}
\affiliation{Max Planck Institute for Chemical Physics of Solids, N\"{o}thnitzer Str.\ 40, D-01187 Dresden, Germany}

\date{\today}
\begin{abstract}

We have studied the temperature-pressure phase diagram of CeCoSi by electrical-resistivity experiments under pressure. Our measurements revealed a very unusual phase diagram. While at low pressures no dramatic changes and only a slight shift of the Ne\'{e}l temperature $T_N$ ($\approx 10$~K) are observed, at about 1.45~GPa a sharp and large anomaly, indicative of the opening of a spin-density-wave (SDW) gap, appears at a comparatively high temperature $T_S \approx 38$~K. With further increasing pressure $T_S$ shifts rapidly to low temperatures and disappears at about 2.15~GPa, likely continuously in a quantum critical point, but without evidence for superconductivity. Even more surprisingly, we observed a clear shift of $T_S$ to higher temperatures upon applying a magnetic field. We discuss two possible origins for $T_S$, either magnetic ordering of Co or a meta-orbital type of transition of Ce.

\end{abstract}

\pacs{}

\maketitle

\section{INTRODUCTION}

Tuning systems from a magnetic to a non-magnetic ground state is presently the most promising way to induce physical properties defying established standard models on solid-state physics, like unconventional superconductivity, unconventional metallic states, or quantum criticality. Intermetallic compounds based on the rare-earth elements Ce or Yb are particularly suited for this strategy. By choosing compounds with appropriate chemical composition and/or applying pressure these elements can easily be tuned from the trivalent state bearing a stable local moment to a tetravalent (Ce) or divalent (Yb) state with an empty (Ce) or a full (Yb) 4$f$ shell, being therefore non-magnetic. In the transition region these systems can be well described within the Kondo-lattice model, where the $f$-moments are dynamically screened by conduction electrons. The related strong quantum fluctuations are likely an essential ingredient for the unconventional properties mentioned above. They can be further enhanced by reducing the dimensionality from 3D to 2D. This has e.g.\ proven to be very effective in stabilizing unconventional superconducting states and enhancing their superconducting transition temperatures.\cite{Petrovic01, Nicklas01} Therefore, compounds crystallizing in the tetragonal CeFeSi structure type,\cite{Bodak70} which can be rationalized as a stacking of a double Ce-layer and a Fe$_2$Si$_2$ layer along the $c$-axis, look promising. This family of compounds is not very numerous, but a few Ce-based compounds have been reported and the ground state of most of them is now well established. Interestingly they cover the whole range from stable trivalent and antiferromagnetically ordered Ce (e.g.\ CeCoGe \cite{Welter94,Chevalier06}) to strongly valence fluctuating and paramagnetic systems (e.g.\ CeFeSi\cite{Welter92}). Examples close to the critical region related to the transition from a magnetic to a non-magnetic ground state are the heavy-fermion systems CeRuSi\cite{Rebelsky88} and CeTiGe\cite{Deppe09}. The latter one presents the strongest metamagnetic transition observed up to now in a paramagnetic Kondo-lattice system \cite{Deppe12}. However, these two compounds are on the non-magnetic side of the critical point and applying pressure is expected to make them even less magnetic. Thus, both compounds are likely rather inappropriate to induce such a critical state under pressure.

On the contrary, CeCoSi was reported to exhibit trivalent Ce ordering antiferromagnetically at 8.8~K \cite{Welter94,Chevalier04,Chevalier06}. The temperature dependence of its resistivity suggests the presence of a quite significant Kondo interaction  \cite{Chevalier04}, locating this compound on the magnetic side of the critical point, but close to it. Thus, a comparatively low pressure might be sufficient to shift this system close to the critical point, and we therefore decided to investigate its physical properties under hydrostatic pressure. When we started this study, only ambient pressure properties, i.e. magnetic susceptibility \cite{Chevalier04, Welter94}, specific heat \cite{Chevalier06}, and electrical resistivity \cite{Chevalier04} had been reported for CeCoSi. We note that recently V.~A. Sidorov \textit{et al.} presented a pressure study at some conferences \cite{Sidorov13}. Their experimental results essentially agree with our observations. However, the residual resistivity ratio of our sample (about 170) is significantly higher than that reported by V.~A. Sidorov \textit{et al.},\cite{Sidorov13} resulting in much sharper signatures at the different transitions. This allows us a more precise determination of the phase diagram. We also performed measurements in magnetic field. The results reveal unexpected and intriguing field dependencies.

\section{Methods}

Polycrystalline samples of CeCoSi  were synthesized by arc melting under argon atmosphere, starting from nominal amounts of the constituents weighted inside a glove-box, (Ce: Ames 99.99\%  (rod), Co: Chempur 99.995\% (wire), Si: Alfa 99.9999\% (lump)). The melted ingots were placed in a tungsten boat wrapped with zirconium foil and annealed for two weeks in two steps, first at 1200$^\circ$C and then at 800$^\circ$C. The quality of the samples  was verified by means of  x-ray powder-diffraction measurements using Cu-K$_{\alpha1}$ radiation ($\lambda=1.54056$~{\AA}) in a Stoe-Stadip-MP diffractometer. All Bragg reflexes could be indexed with the CeFeSi structure type and the observed lattice parameters, $a = 0.4054$~nm and $c = 0.6978$~nm, are close to those reported in literature.\cite{Welter94,Chevalier04} The specific heat and the magnetic susceptibility were measured using a Physical Property Measurement System (PPMS) and a Magnetic Property Measurement System (MPMS), respectively, both from Quantum Design. A double-layered piston-cylinder type pressure cell has been used for generating pressures up to 3~GPa. In order to achieve hydrostatic pressure conditions, silicon-oil served as pressure transmitting medium. The pressure was monitored at low temperatures, by the shift of the superconducting transition temperature of a piece of Pb placed inside the pressure chamber. From the width of the superconducting transition, a pressure gradient smaller than 0.027~GPa could be estimated for the entire pressure range. The electrical-resistivity measurements have been performed by using a four-terminal a.c.\ technique. A Linear Research LR700 resistance bridge and a Lock-in amplifier together with a low-temperature transformer have been utilized for the measurements. Temperatures down to 40~mK have been obtained in a Kelvinox 100 dilution refrigerator from Oxford Instruments, while in the temperature range $1.8~{\rm K}\leq T\leq 300~{\rm K}$ a PPMS has been employed. The experiments were carried out in magnetic fields up to 9~T.

\begin{figure}[b!]
\includegraphics[angle=0,width=8cm,clip]{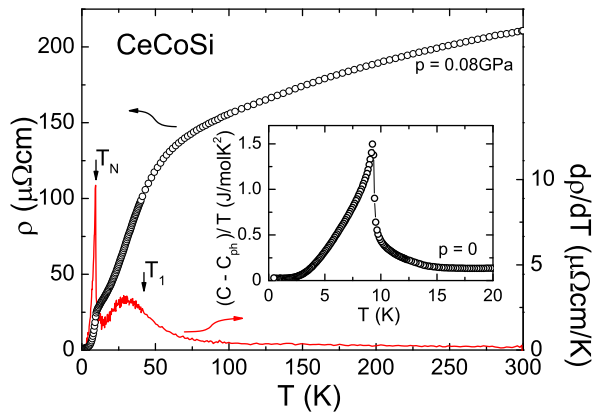}
\caption{\label{rhoCp} (Color online) Temperature dependence of the electrical resistivity ($\rho$) of CeCoSi (left axis) and of ${\rm d}\rho/{\rm d} T$ (right axis). $T_N$ and $T_1$ are indicated by arrows. The inset displays the specific heat of CeCoSi as $(C-C_{ph})(T)/T$. $C_{ph}$ has been determined from the specific heat of LaCoSi.}
\end{figure}

\section{EXPERIMENTAL RESULTS}

\subsection{Ambient pressure}

\begin{figure}[tb!]
\includegraphics[angle=0,width=8cm,clip]{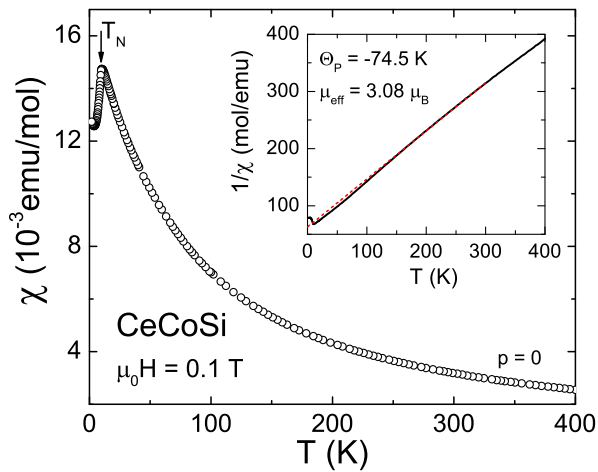}
\caption{\label{Chi} (Color online) Magnetic susceptibility, $\chi(T)=M(T)/H$, of CeCoSi taken in a magnetic field of $\mu_0H=0.1$~T. The arrow marks $T_N$. The inset shows the inverse susceptibility $1/\chi(T)$. The paramagnetic Curie temperature, $\theta_P$, and the effective magnetic moment, $\mu_{\rm eff}$, are obtained from a fit of a Curie-Weiss law $\chi(T)=C/(T-\theta_P)$ in the temperature interval $150~{\rm K} \leq T \leq 300~{\rm K}$ (red dashed line in the inset).}
\end{figure}

The electrical resistivity, $\rho(T)$, measured at ambient pressure is displayed in Fig.~\ref{rhoCp}. A very good quality of our polycrystalline samples is suggested by the high residual resistivity ratio (RRR) of about 170. The temperature dependence of the electrical resistivity is reminiscent of a Kondo-lattice system. $\rho(T)$ shows a metallic behavior at high temperatures followed by the onset of coherent Kondo scattering below 70~K. At low temperatures, a clear kink is observed at $T_N \approx 9.5$~K, corresponding to the phase transition into the antiferromagnetically ordered state. Our value obtained for $T_N$ is slightly higher than the value reported previously \cite{Chevalier04}. The temperature derivative of $\rho(T)$, ${\rm d}\rho(T)/{\rm d} T$, shows a $\lambda$-like peak at $T_N$ (see Fig.~\ref{rhoCp}). Additionally, a relatively broad maximum is seen at higher temperatures. This feature is characterized by a broad minimum in the second temperature derivative of $\rho(T)$, ${\rm d^2}\rho(T)/{\rm d} T^2$, around $T_1 \approx 42.5$~K. The magnetic-susceptibility data, shown in Fig.~\ref{Chi}, exhibits a Curie-Weiss-like temperature dependence at high temperatures, with a clear signature of the antiferromagnetic (AFM) phase transition at lower temperatures. A linear fit to the $1/\chi(T)$ above 150~K yields a paramagnetic Curie temperature $\theta_P \approx -74.5$~K and an effective magnetic moment $\mu_{\rm eff} \approx 3.08~\mu_B$, in reasonably good agreement with the values reported in the literature ($\mu_{\rm eff}\approx2.71~\mu_B$/Ce and $\theta_P\approx -55$~K) \cite{Chevalier04,Welter94}. It is worth mentioning that the value of $\mu_{\rm eff}$ is either way larger than the $2.54~\mu_B$ expected for the free Ce$^{3+}$ ion. This might give a hint at a magnetic moment on the Co site. Finally, the specific-heat data (inset of Fig.~\ref{rhoCp}) show a clear anomaly at low temperatures corresponding to the entrance into the antiferromagnetically ordered state. The large value of the specific heat at $T_N$ of 13.5~J/(mol K) after subtracting the phonon contribution determined from the specific heat of LaCoSi ($\gamma\approx0.0145$~J/(molK$^2$)), which exceeds the value 12.5~J/(mol K) expected for a mean value $S=1/2$ system, as well as the large amount of 4$f$ entropy collected just above $T_N$ as $S({\rm 10~K})=0.74~{\rm R\ln2}$ indicate that the Kondo interaction is much weaker than the RKKY interaction. This contrasts the pronounced Kondo-like behavior observed in $\rho(T)$ below 70~K. In view of further results presented below, this calls for a different explanation for the pronounced decrease of $\rho(T)$ below 70~K. This will be discussed in Section~\ref{discussion}.

\subsection{Hydrostatic pressure}

\begin{figure}[tb!]
\includegraphics[angle=0,width=8cm,clip]{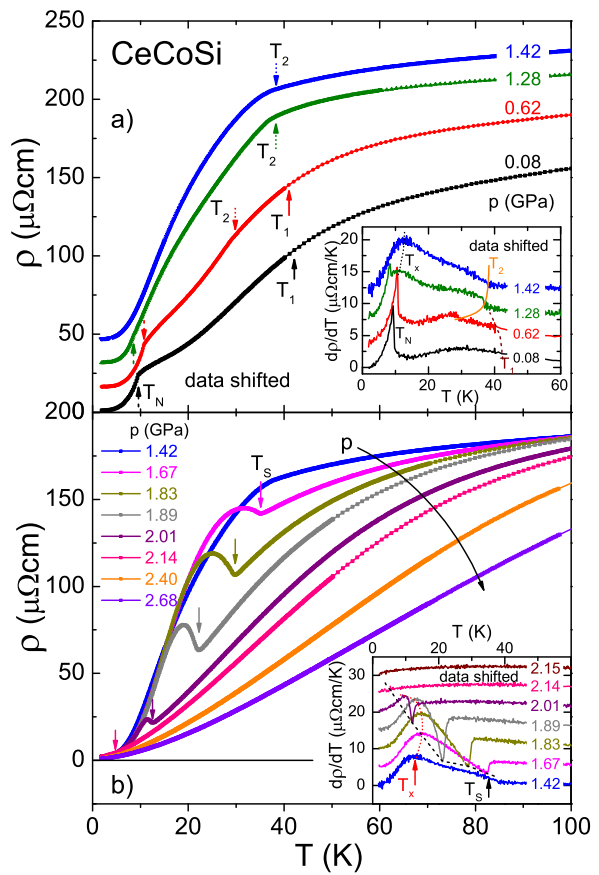}
\caption{\label{rho_overview} (Color online) Electrical resistivity, $\rho$, of CeCoSi as function of temperature for different pressures. Arrows mark the transition, respectively, crossover temperatures as indicated in the figure. $\rho(T)$ at low pressures ($p\leq1.42$~GPa) is shown in (a). The data sets have been shifted by $15~\mu\Omega$cm with respect to each other for clarity. The resistivity data for $p\geq1.42$~GPa are displayed in (b). The insets show the corresponding temperature derivatives, ${\rm d}\rho(T)/{\rm d} T$. The curves in both insets have been shifted for clarity.}
\end{figure}

We now turn to the resistivity study on CeCoSi under hydrostatic pressure. The results in the temperature range up to 100~K are shown in Fig.~\ref{rho_overview}. At room temperature, we find a smooth evolution of $\rho_{\rm 300\,K}$ with pressure. $\rho_{\rm 300\,K}(p)$ stays almost constant. The $\rho(T)$ data show different anomalies which exhibit distinct pressure dependencies. The phase-transition anomaly observed at $T_N\approx9.5$~K at atmospheric pressure initially shifts to higher temperatures with increasing pressure, reaches a maximum temperature of $T_N\approx10.7$~K around 0.62~GPa and starts to move to lower temperatures again. $T_N(p)$ disappears in a first-order-like way between 1.28 and 1.42~GPa. The broad minimum in ${\rm d^2}\rho(T)/{\rm d}T^2$ at $T_1$ found at ambient pressure shifts slightly to lower temperatures upon increasing pressure and gets absorbed in a new anomaly $T_2$ at $p\approx 1.28$~GPa. This might suggest some connection to $T_N$, whose signature gets lost in the same pressure region. However, at ambient pressure, the magnetic susceptibility does not show any feature at $T_1$ suggesting that $T_1$ marks only a crossover between different temperature regimes.

\begin{figure}[tb!]
\includegraphics[angle=0,width=8cm,clip]{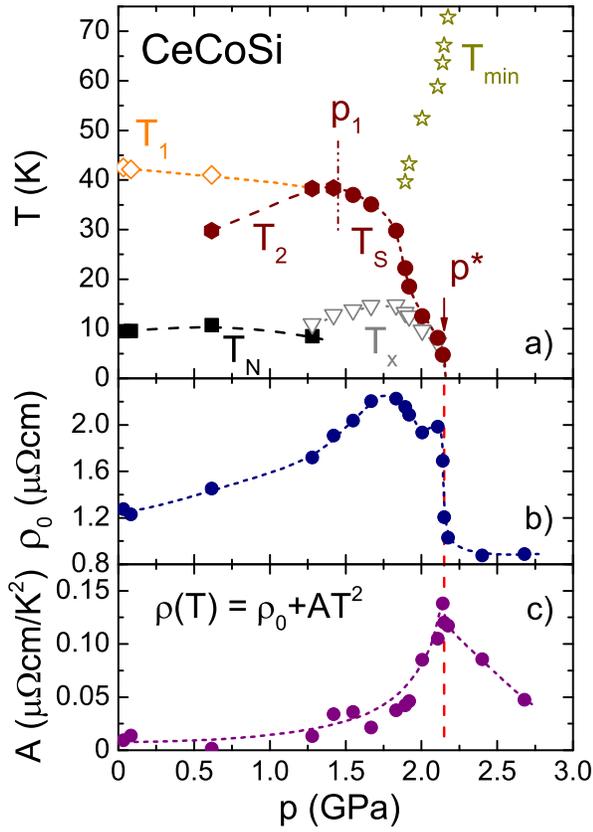}
\caption{\label{CeCoSi_PhD} (Color online) Temperature-pressure ($T-p$) phase diagram of CeCoSi. Full symbols correspond to phase transitions, while open symbols represent crossover temperatures. $p_1$ indicates a characteristic change of the behavior of the electrical resistivity as function of pressure, see text for details. $p^*$ marks the pressure where the magnetic order at $T_S$ is suppressed to zero temperature. Pressure dependence of the residual resistivity $\rho_0$ (b) and of the $A$ coefficient (c) of a $\rho(T)=\rho_0+AT^2$ fit to the low-temperature data (see text for details).}
\end{figure}

Under pressure, additional anomalies and features appear in the electrical resistivity. We first concentrate on the characteristics marked with $T_2$ and $T_x$ in Fig.~\ref{rho_overview}. At 1.28~GPa, the last pressure where we can identify $T_N$ in $\rho(T)$, a second feature appears at $T_x$ slightly above $T_N$. We define $T_x$ by the position of the maximum in ${\rm d}\rho(T)/{\rm d}T$. The feature at $T_x$ seems to be closely related to the magnetic ordering at $T_N$. This is supported by measurements in applied magnetic fields. A detailed discussion of the data obtained in magnetic fields follows later on. Upon increasing pressure, $T_x(p)$ exhibits a maximum around 1.75~GPa before it is suppressed to $T=0$ close to 2.15 GPa (see Fig.~\ref{CeCoSi_PhD}a).

Furthermore, at 0.62~GPa a kink starts to develop in $\rho(T)$ denoted as $T_2$ in Fig.~\ref{rho_overview}a. This kink corresponds to a step-like feature in ${\rm d}\rho(T)/{\rm d}T$, which moves to higher temperatures upon increasing pressure ($p \leq 1.42$~GPa). Above 1.42~GPa this kink transforms into a cusp-like anomaly in the resistivity. We mark this pressure as $p_1$. The observed shape in $\rho(T)$ is characteristic for the opening of a gap at the Fermi-level, e.g. due to the formation of a spin-density-wave (SDW) type of magnetic ordering below $T_S$. Between 1.67 and 1.89 GPa the anomaly at $T_S$ becomes more pronounced with increasing pressure. We do not observe any thermal hysteresis in $\rho(T)$ at $T_S$. This indicates a second-order type phase transition. Upon increasing pressure $T_S(p)$ is rapidly suppressed. We lose the signature of the phase transition in a very narrow pressure range. At 2.14~GPa we still find a transition temperature of $T_S=4.77$~K, while at $p = 2.15$~GPa no phase transition anomaly can be identified anymore down to 40~mK. We note that $T_x(p)$ and $T_S(p)$ disappear at the same pressure $p^*\approx 2.15$~GPa.

From this pressure on, the electrical resistivity shows a metallic temperature dependence down to the lowest temperatures of our measurements. The only remaining feature in the resistivity in our experimental pressure range ($p\lesssim2.7$~GPa) is a broad minimum in ${\rm d^2}\rho(T)/{\rm d}T^2$, indicated as $T_{\rm min}$. It shifts to higher temperatures upon increasing pressure and might be related to a crossover between different temperature regimes.

\subsection{Effect of magnetic field}

\begin{figure}[tb!]
\includegraphics[angle=0,width=8cm,clip]{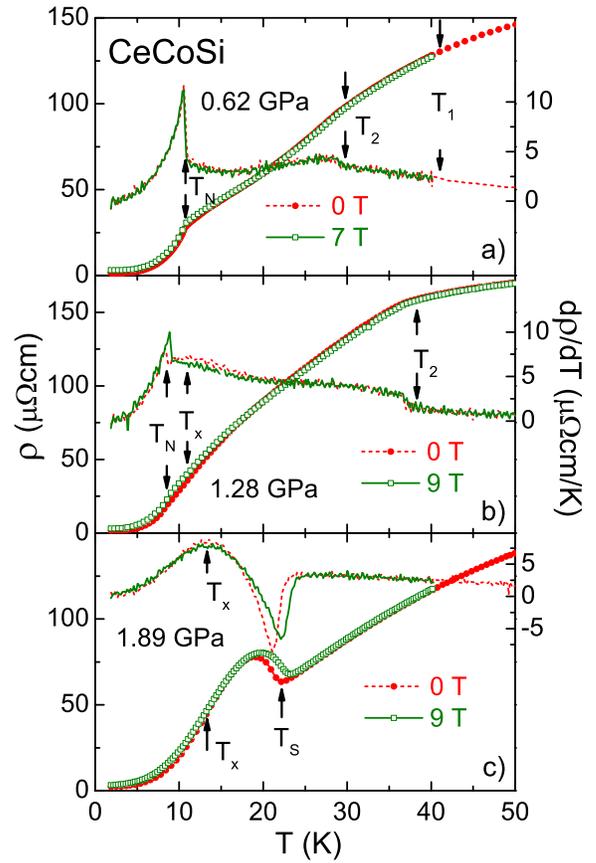}
\caption{\label{rho_field} (Color online) Temperature dependent electrical resistivity at 0.62 (a), 1.28 (b), and 1.89~GPa (c) in magnetic fields of 0 and 9(7)~T (left axis) and the corresponding temperature derivatives ${\rm d}\rho(T)/{\rm d} T$ (right axis). The arrows indicate the transition/crossover temperatures in $H=0$ (see text for details).}
\end{figure}

We have also investigated the effect of a magnetic field on the electrical resistivity of CeCoSi. We note that the measurements have been performed on polycrystalline samples. Figure~\ref{rho_field} shows $\rho(T)$ and ${\rm d}\rho(T)/{\rm d}T$ for $\mu_0H = 0$ and 9(7)~T for the representative pressures 0.62, 1.28, and 1.89~GPa. At $p\leq0.62$~GPa, $T_N$ indicated by the sharp peak in ${\rm d}\rho(T)/{\rm d} T$ exhibits a tiny shift to lower temperatures upon increasing the magnetic field. A very weak field dependence of $T_N$ was already observed by Chevalier \textit{et al.} \cite{Chevalier06b}. Furthermore, the size of the peak is suppressed by the magnetic field. The opposite is observed at 1.28~GPa. Here, the peak in the temperature derivative of $\rho(T)$ marking $T_N$ shifts significantly to higher temperatures and becomes much more pronounced in magnetic field. Interestingly, the feature at $T_x$ becomes washed out in 9~T, which might indicate some relation between $T_N$ and $T_x$. On the other two features at $T_1$ and $T_2$ a magnetic field of $\mu_0H=9$~T has only a minor effect. At 1.89~GPa, which is representative for the pressure region between 1.45 and 2.15~GPa, ${\rm d}\rho(T)/{\rm d} T$ shows a maximum at $T_x$, which is only little affected by a magnetic field. We note that $\rho(T)$ exhibits a small positive magnetoresistance for $T<T_x$ (for $\mu_0H\leqslant9$~T).

The strongest effect of the magnetic field is observed on the transition at $T_S$. At all pressures $T_S(H)$ shifts significantly to higher temperatures upon increasing the magnetic field, while the shape of the anomaly in $\rho(T)$ at $T_S$ remains essentially unchanged. This dependence of $T_S$ in magnetic field is rather unexpected and will be discussed in details below.

\section{DISCUSSION}\label{discussion}

The temperature-pressure phase diagram of CeCoSi, shown in Fig.~\ref{CeCoSi_PhD}a, deduced from our resistivity data presents one very unusual and intriguing feature: the quite abrupt appearance of a transition at a much higher temperature than those observed at lower pressures, with a signature implying the opening of a gap at the Fermi level, but with a significant shift to higher temperatures under magnetic field. This field-induced shift implies that this transition is closely related to a magnetic state. A valence transition exhibits a magnetic field dependence too, but in the opposite direction: in valence-fluctuating systems the high-temperature state corresponds to the magnetic state, which gets stabilized by an external magnetic field. Therefore, a valence transition shifts to lower temperatures with increasing magnetic field, and can thus be excluded in the present case. The opening of a gap at the Fermi level evidenced by the upturn in $\rho(T)$ would imply a SDW type antiferromagnetic order. In this case one also expects a decrease of the ordering temperature in an external magnetic field, in disagreement with our observations on CeCoSi. A shift of the anomaly to higher temperatures in a magnetic field would be in accordance with a ferromagnetic state, but in this case a rapid broadening and weakening of the anomaly with increasing the magnetic field is anticipated, since a magnetic field changes a second-order ferromagnetic transition into a crossover. Furthermore, in magnetic rare-earth intermetallic systems the opening of a gap is not observed at a ferromagnetic transition. Thus, none of the standard scenarios can explain the observed behavior.

A first hint towards possible scenarios is suggested by $T_S$ emerging from the shoulder in $\rho(T)$ (i.e.\ the maximum in the curvature) at $T_1$  at low pressures. Such a maximum can be attributed to two quite different origins. On the one hand, a pronounced negative curvature in $\rho(T)$ below 200 K is typical for transition metal spin-fluctuating systems. This would imply Co to be close to a magnetic state and becoming truly magnetically ordered above a pressure of 1.45~GPa. Support for this scenario is given by the larger than expected $\mu_{\rm eff}$ observed at $p = 0$. Furthermore, a very weak magnetic order of Co was recently reported in LaCoGe based on NMR spectra, but no anomaly related to the magnetic transition was observed in $\rho(T)$ \cite{Karube13}. The onset of Co magnetism would trivially account for a large increase in the ordering temperature. However, a Pauli-like susceptibility has been reported for LaCoSi \cite{Welter94}. Moreover, itinerant Co magnetism is typically suppressed under hydrostatic pressure. Accordingly, the substitution of Si for Ge in LaCoGe, corresponding to a chemical pressure of the order of 10~GPa, is expected to suppress the magnetic order of Co in LaCoSi completely, consistent with the experimental observations \cite{Welter94}. Neither would pressure be expected to induce a magnetically ordered Co state in CeCoSi. Furthermore, it is difficult to conceive why pressure first induces magnetic Co ordering, and then at just a slightly larger pressure magnetic order disappears again. In addition, Co magnetism does not provide a simple way to understand the discrepancy between the gap opening and the positive ${\rm d}T_S(H)/{\rm d}H$. Thus, while the onset of Co magnetism is a possible scenario to explain the appearance of a transition at a comparatively large temperature (for Ce-systems), it does not explain all experimental observations consistently. A related scenario would be that Ce polarizes Co in CeCoSi, resulting in a common ordering of Ce and Co at $T_S$. However, this scenario presents the same problems as the scenario with pure Co ordering discussed above, and is, therefore, also unlikely.

On the other hand, in Kondo-lattice systems a shoulder in $\rho(T)$ is quite common and attributed to the Kondo scattering by an excited crystal electric field (CEF) level [Note:\ one can exclude this feature to be caused by Kondo scattering by the ground state CEF level because the 4$f$ entropy reaching $0.5~{\rm R\ln2}$ at 8.6~K implies a Kondo temperature below 17~K]. The shift of this shoulder at $T_1$ to lower temperatures with increasing pressure indicates a decrease of the CEF splitting. At some point, when the Zeeman splitting connected with the exchange field becomes of the order of the CEF splitting, the exchange field should be able to involve the excited CEF level into the magnetic order. This might occur through level mixing (smooth transition) or level crossing (sharp transition). Since the excited level might have a larger moment and/or a larger exchange than the CEF ground state, this can result in a significant increase of the ordering temperature. The switch from $T_N$ to $T_S$ would then correspond to the magnetic version of the "meta-orbital" transition recently invoked for explaining the properties of CeCu$_2$Si$_2$ under pressure.\cite{Hattori10,Pourovskii13} Furthermore, because the system then corresponds to a quasi-quartet, even quadrupolar order might develop, which is quite often associated with an increase of the transition temperature with magnetic field.\cite{Effanti85} Recently, such a transition from an AFM to a quadrupolar state upon applying pressure has been observed in CeTe.\cite{Kawarasaki11} In turn, becoming a quasi-quartet implies a doubling of the effective ground state degeneracy, which is expected to strongly enhance the Kondo temperature. This would explain the rapid decrease of $T_S$ and the disappearance of the magnetic order upon further increasing the pressure. Thus, while solely resistivity data do not allow for a determination of the nature of the transition at $T_S$, quite a number of arguments favor the involvement of the CEF excited state in the appearance of this transition. This discussion leaves the nature of the transition at $T_2$ and of the anomaly at $T_x$ open. The pressure dependence of $T_2$ strongly suggests that this transition is a precursor of $T_S$, likely with a strongly reduced order parameter resulting in a much weaker anomaly. The shape of the anomaly at $T_x$, i.e. a broad bump just above $T_N$ at $p = 1.28$~GPa, and the fact that at this pressure the feature weakens under magnetic field while the anomaly at $T_N$ becomes stronger indicate that the anomaly at $T_x$ reflects short range correlations, likely quasi-two-dimensional in nature, connected with $T_N$. This suggests a competition between two quite different order parameters, one dipolar magnetic associated with $T_N$ and one dipolar magnetic or quadrupolar associated with $T_S$.

Two characteristic pressures, $p_1 = 1.45$~GPa and $p^* = 2.15$~GPa, can be defined. In the following, we want to explore the possibility of the existence of a quantum critical point (QCP) at $p_1$ and/or $p^*$, respectively. In many Ce-based heavy-fermion materials a superconducting phase emerges in close proximity to a QCP. To probe for superconductivity, we carried out $\rho(T)$ experiments down to 40~mK at 1.42~GPa close to $p_1$ and at 2.15~GPa at $p^*$. At both critical pressures we did not find a transition to a zero resistance state, excluding superconductivity above 40~mK. To further characterize the ground state of CeCoSi we analyzed the low-temperature resistivity data. In the entire pressure range of our study ($p\lesssim2.7$~GPa) the low-temperature $\rho(T)$ data can be fit by $\rho(T)=\rho_0+AT^2$. This temperature dependence is predicted for a Landau-Fermi-liquid (LFL). The pressure dependencies of $\rho_0$ and $A$ obtained by fitting the $\rho(T)$ data in the temperature range $1.8~{\rm K} \leq T \leq 5$~K are depicted in Fig.~\ref{CeCoSi_PhD}b and \ref{CeCoSi_PhD}c. Enhanced values of $\rho_0(p)$ and $A(p)$ can be found around $p^*$, suggesting the presence of strong magnetic fluctuations in this pressure region. $\rho_0(p)$ shows a broad maximum at $p \approx 1.75$~GPa. We can only speculate, whether this maximum in $\rho_0(p)$ is related to the maximum in $T_x(p)$ and/or to the start of the abrupt suppression of $T_S(p)$ with increasing pressure.

\section{SUMMARY}

To conclude, we have performed electrical-resistivity measurements under hydrostatic pressure ($p \lesssim 2.7$~GPa) on the layered Kondo-lattice system CeCoSi and established a complex magnetic $T-p$ phase diagram, with multiple phase transitions. The AFM ordering of the Ce-sublattice observed at $T_N \approx 9.5$~K at ambient pressure shifts first slightly up and then down in temperature with increasing pressure and disappears quite abruptly around $p_1= 1.45$. This ordering seems to be replaced by short-range fluctuations as suggested by a broad maximum in the temperature derivative of $\rho(T)$ at $T_x \approx 15$~K. However, at a lower pressure of about 0.62~GPa a new transition appears at $T_2 = 29.8$~K, shifts to higher temperatures with increasing pressure, and merges at around 1.28~GPa and $T \approx 38$~K with the onset of the pronounced decrease in $\rho(T)$ already present at $p = 0$, which is usually attributed to onset of coherence in a Kondo lattice. At the slightly larger pressure $p_1$, this transition strongly changes its character, showing now a well pronounced upturn in $\rho(T)$ usually associated with the opening of a gap at the Fermi surface connected with a SDW transition. This indicates a significant change in the order parameter, thus we named this transition $T_S$ to mark the difference to $T_2$. Since the onset of $T_S$ and the disappearance of $T_N$ occurs simultaneously, this new order parameter seems to compete with the ambient pressure AFM order. With further increasing pressure $T_S$ decreases more and more rapidly, absorbs $T_x$, and finally drops toward $T = 0$ at $p^*=2.15$~GPa. Although the slope ${\rm d}T_S(p)/{\rm d}p$ is quite large, it seems that the ordered state disappears in a continuous way and thus results in a QCP at $p^*$. However, at low temperatures $\rho(T)$ obeys a Fermi-liquid behavior at all pressures, merely showing a clear peak in the quasiparticle-quasiparticle scattering at $p^*$. Furthermore, the residual resistivity, which increases quite continuously with pressure at low pressures and shows a maximum between $p_1$ and $p^*$, also drops abruptly at $p^*$. Beyond $p^*$ $\rho(T)$ exhibits a simple metallic behavior without any evidences for a phase transition. We did not find any indication for superconductivity in the whole $T - p$ phase diagram. Thus, while all features at $p^*$ indicate a continuous switch from an ordered, but strongly fluctuating state to a normal metallic state, some of these features do not follow the standard expectation for a QCP. The most unusual feature in this phase diagram is the quite abrupt appearance of a SDW transition at $p_1$, at a quite large temperature, a factor of 4 larger than $T_N$ at ambient pressure. Surprisingly, this transition shifts to higher temperatures in a magnetic field, in strong contrast to the expectation for a SDW. This transition appears at the location of the shoulder in $\rho(T)$ associated with Kondo scattering by the first excited crystal field level. Therefore, we propose that this transition is connected with the involvement of the excited CEF doublet in the formation of the ordered state, caused by a decrease of the CEF splitting and an increase of the exchange field upon applying pressure. The resulting effective quasi-quartet would allow for quadrupolar ordering, which opens a way to reconcile a SDW type anomaly with a positive slope ${\rm d}T_S(H)/{\rm d}H$. However, with only resistivity data we cannot exclude a very different scenario in which $T_S$ is connected with magnetic order of Co. But many of the observed features are difficult to explain within such a scenario.

\section*{ACKNOWLEDGMENTS}
We acknowledge fruitful discussions with J.\ G.\ Sereni and thank T.\ Gruner and R.\ Koban for experimental support.

\end{document}